\documentclass[a4paper,10pt]{article}
\usepackage[utf8]{inputenc}

\usepackage{publication}
\usepackage[backend=biber,sorting=none]{biblatex}
\usepackage{amsmath} 
\usepackage{graphicx}
\usepackage{algorithm}
\usepackage{algpseudocode}
\usepackage[misc]{ifsym}

\DeclareMathOperator \supp{supp}

\title{Reducing ringing artifacts in Wiener deconvolution using implicit physical priors}

\author[1,\Letter]{Jakub Czuchnowski}
\contact{\textsuperscript{\Letter}Correspodning author: Jakub Czuchnowski - jczuchno@bu.edu}

\author[1,2]{Jerome Mertz}

\affil[1]{Department of Biomedical Engineering, Boston University, Boston, MA, 02215, USA}
\affil[2]{Photonics Center, Boston University, Boston, MA, 02215, USA}

\date{\today}

\begin{document}

\maketitle

\begin{abstract}
    With advances in high-speed imaging there is a growing need for robust high-speed deconvolution algorithms. Wiener deconvolution remains one of the fastest and simplest algorithms available, however it suffers from ringing artifacts that can limit its applicability. In this work, we develop two complementary, computationally efficient heuristic methods to suppress  these ringing artifacts by using physical non-negativity and sparsity priors. 
\end{abstract}



\begin{multicols}{2}

\section{Introduction}

The performance of imaging systems is often limited by optical aberrations, which reduce both image resolution and signal-to-noise ratio (SNR). As a result, considerable effort has gone into the development of methods to correct these aberrations, either by sophisticated lens design \cite{kingslake2009lens} or by incorporating active components \cite{booth2007adaptive,hampson2021adaptive} that compensate for distortions during image acquisition. However, because advanced correction techniques often increase cost and system complexity, deconvolution remains the most widely used approach for aberration correction.

Among the many available deconvolution methods, Wiener deconvolution remains one of the most popular because of its computational simplicity and high speed. However, typical problems that arise when using Wiener deconvolution are the so called 'ringing' artifacts that manifest themselves as halos around sparse objects or corrugated backgrounds in case of denser samples. While methods have been developed to mitigate these artifacts, they typically require extensive processing or iterative optimization \cite{dimlo2023improved, mosleh2017explicit, vsroubek2019iterative,zhao2020natural}, which compromises the high-speed advantage of Wiener deconvolution. Here, we develop two complementary, computationally efficient heuristic methods to address these ringing artifacts by using physical priors of non-negativity and sparsity. 

\section{Theory}

Wiener deconvolution is a method based on reweighting the Fourier spectrum of an image:
\begin{equation}
    W(I|\epsilon)=FT^{-1}\bigg\{\frac{\hat{I}\cdot\hat{h}^*}{|\hat{h}|^2+\epsilon} \bigg\}
\end{equation}
where $\hat{I}$ is the Fourier transform of the image ($I$), $\hat{h}^*$ in the complex conjugate of the Fourier transform of the point spread function (PSF, $h$) and $\epsilon$ is a regularization parameter. 

As noted above, a typical problem with using Wiener deconvolution is the generation of ringing artifacts. These originate from the modified optical transfer function which for a Wiener filter approaches a top-hat, making the effective PSF oscillatory. We can thus write
\begin{equation}
    W(I|\epsilon)\approx o+r_N+r_P
\end{equation}
where $r_N$ and $r_P$ are the respectively the negatively and positively valued ringing artifacts. We describe two approaches to suppress these artifacts.

\subsection{Multiscale Positivity Constraint (MPC)}

\begin{figure}[H]
\begin{center}
\includegraphics[width=\linewidth]{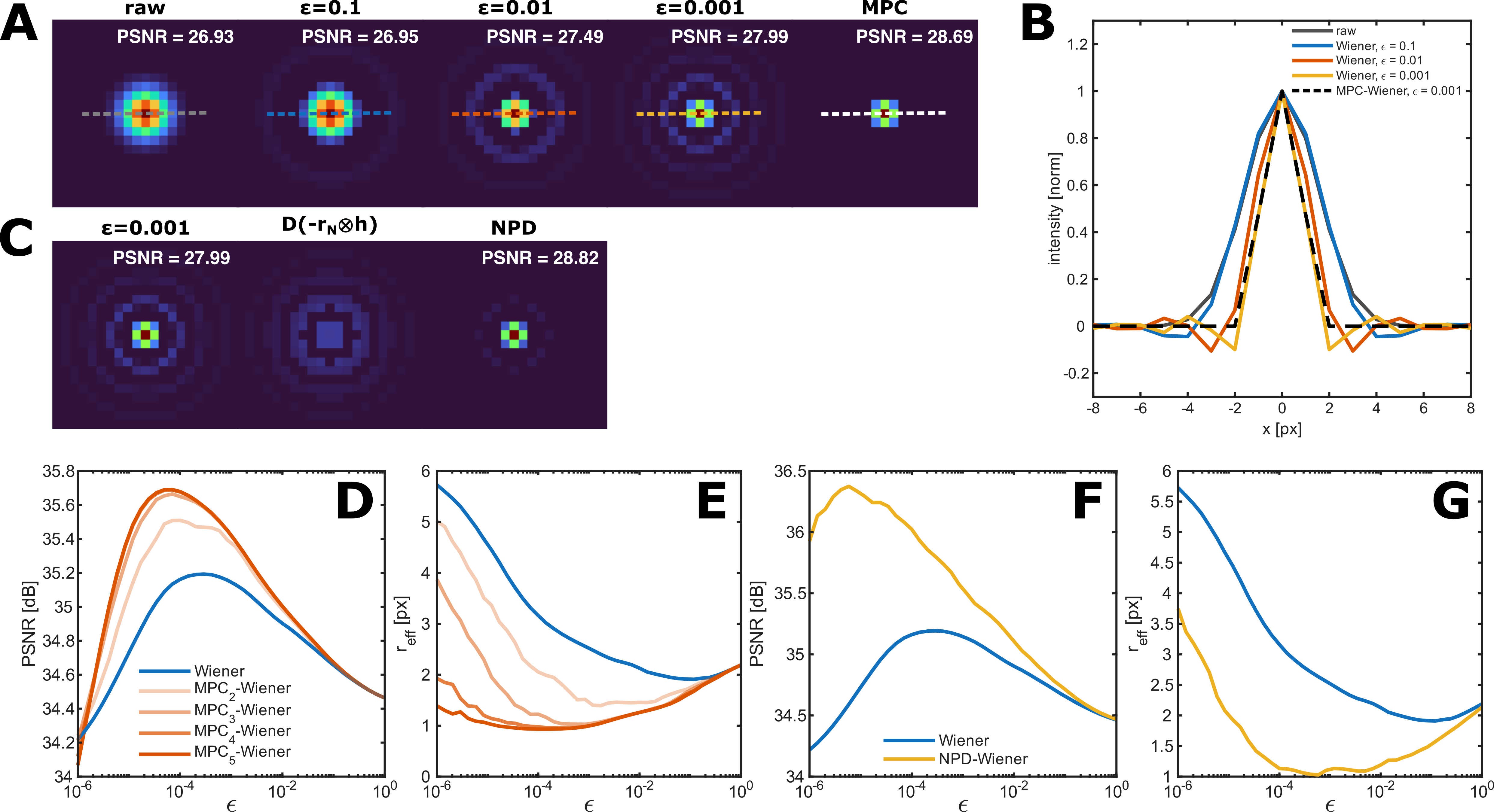} 
\caption 
{\label{fig:1}
\textbf{(A)} Illustration of the positive parts of PSFs Wiener-deconvolved using different values of $\epsilon$. \textbf{(B)} Line-plots across the PSFs from (A) showing how MPC negates ringing artifacts. \textbf{(C)} Principle of NPD operation which calculates and subtracts $D(-r_N\otimes h)$. \textbf{(D,F)} Dependence of the PSF's PSNR on the regularization constant $\epsilon$ for conventional Wiener compared to MPC-Wiener \textbf{(D)} and NPD-Wiener \textbf{(F)}. \textbf{(E,G)} Dependence of the effective radius of the PSF on the regularization constant $\epsilon$ for conventional Wiener compared to MPC-Wiener \textbf{(E)} and NPD-Wiener \textbf{(G)}. For \textbf{(D-G)} The PSF was adjusted to a peak intensity of $10^3 \ e^-$ and appropriate shot noise was added before deconvolution.}
\end{center}
\end{figure}

In a first approach, the negative-valued ringing artifacts ($r_N$) are interpreted as not carrying physical information, meaning they can be readily removed by imposing a positivity constraint \cite{luo2019deblurring} that assumes there is a low likelihood of an object where the Wiener-filtered image has negative values. That is,
\begin{equation}
    [W(I|\epsilon)](x,y)<0 \implies o(x,y)=0 
    \label{eq:positivity}
\end{equation}
which leads to 
\begin{equation}
    W(I|\epsilon)_{PC}=
    \left\{\begin{matrix}
       W(I|\epsilon;x,y) & \text{for } [W(I|\epsilon)](x,y)\geq0 \\
       0 & \text{for } [W(I|\epsilon)](x,y)<0
    \end{matrix}\right.
\end{equation}
where $W(I|\epsilon)_{PC}$ is the positive-constrained Wiener image. However, this method cannot remove the positive-valued ringing artifacts ($r_P$). To address this, we propose to generalize the assumption of Eq. \ref{eq:positivity} to include \emph{all} values of $\epsilon'>\epsilon$
\begin{equation}
    \exists_{\epsilon'\geq \epsilon}: [W(I|\epsilon')](x,y)<0 \implies o(x,y)=0
\end{equation}
resulting in the generalized positivity constraint (\textbf{Equation \ref{eq:W_MPC}}) where $W(I|\epsilon)_{MPC}$ is the multiscale-positivity-constrained Wiener image. The rationale behind this approach lies in the fact that increasing $\epsilon$ increasingly suppresses higher spatial frequencies that might be corrupted by noise, causing deconvolved images to be less prone to artifacts. Hence, negative values present in the deconvolved images with a higher values $\epsilon$ should be increasingly less likely to coincide with parts of the object. Additionally, since the periodicity of the ringing artifacts depends on $\epsilon$ (\textbf{Figure \ref{fig:1}A}), extending the positivity constraint to Wiener deconvolutions with different $\epsilon$'s efficiently removes ringing (\textbf{Figure \ref{fig:1}B}).

\end{multicols}

\noindent\makebox[\linewidth]{\rule{\linewidth}{0.4pt}}
\begin{equation}
    W(I|\epsilon)_{MPC}=
    \left\{\begin{matrix}
       [W(I|\epsilon)](x,y) & \text{for } \forall_{\epsilon'\geq \epsilon}: [W(I|\epsilon')](x,y)\geq0 \\
       0 & \text{for } \exists_{\epsilon'\geq \epsilon}: [W(I|\epsilon')](x,y)<0
    \end{matrix}\right.
    \label{eq:W_MPC}
\end{equation}
\noindent\makebox[\linewidth]{\rule{\linewidth}{0.4pt}}

\begin{multicols}{2}

Because the evaluation of $W(I|\epsilon)$ for every $\epsilon'>\epsilon$ is intractable in practice, we tested this approach using a variable number ($n$) of discrete $\epsilon'$ values logarithmically spaced within the interval $[\epsilon,1]$. To quantify deviations from ground truth ($GT$), we adopt the metric PSNR, defined by \begin{equation}
    \text{PSNR}(I,GT)=10\log_{10}\bigg(\frac{\max(GT)^2}{\text{MSE}(I,GT)}\bigg)
    \label{eq:PSNR}
\end{equation}
where $\text{MSE}(I,GT)=\sum_i(I_i-GT_i)^2/N_{px}$ is the mean-squared-error and $N_{px}$ is the number of pixels. From \textbf{Figure \ref{fig:1}D} we observe that even $n=2$ provides a noticeable improvement in PSNR, with gains increasing for larger $n$ though with diminishing returns (especially for $n>4$). Additionally, we quantified  the effect of MPC-Wiener on the effective radius ($r_{eff}$) of the deconvolved PSF, which is a measure of the PSF compactness 
\begin{equation}
    r_{eff}(I)=\frac{\sum_{i}I_{i}\sqrt{(x_i-x_c)^2+(y_i-y_c)^2}}{\sum_{i}I_{i}}
\end{equation}
where $x_c=\sum_i x_iI_i/\sum_iI_i$, $y_c=\sum_i y_iI_i/\sum_iI_i$. We observe that standard Wiener deconvolution struggles to reduce the effective radius of the PSF (\textbf{Figure \ref{fig:1}E}) becasue of ringing artifacts, whereas MPC-Wiener significantly reduces $r_{eff}$ while at the same time improving PSNR.

\subsection{Negativity Pseudo-Deconvolution (NPD)}

Our second approach capitalizes on the observation that even though the deconvolved image contains ringing artifacts:
\begin{equation}
    W(I|\epsilon)\approx o+r_N+r_P
\end{equation}
a re-convolution of this with the original PSF allows us to recover the original image (under the assumption that $\epsilon$ is small). That is,
\begin{equation}
    (o+r_N+r_P)\otimes h \approx I=o\otimes h
\end{equation}

This suggests that ringing artifacts vanish after re-convolution:
\begin{equation}
    (r_N+r_P) \otimes h \approx0 \implies r_P\otimes h \approx -r_N\otimes h,
\end{equation}
which links positive and negative ringing artifacts:
\begin{equation}
    r_P\approx D(-r_N\otimes h),
\end{equation}
where $D(\cdot)$ denotes a deconvolution operation. However, we note that the spatial supports of $r_P$ and $r_N$ are disjointed ($\supp \{r_N\} \cap \supp \{r_P\}=\emptyset$), meaning that $D(\cdot)$ needs to be a custom deconvolution method. While this can be achieved using matrix-inversion-based methods, the resulting computational overhead rivals that of iterative deconvolution methods to the point that Wiener deconvolution no longer provides a speed advantage. We thus use a heuristic approximation of $D(\cdot)$ instead, which we call a pseudo-deconvolution (see \textbf{Algorithm \ref{al:1}})
\begin{equation}
    D(-r_N\otimes h) \approx \text{PD}[W(I|\epsilon)]
\end{equation}
which can estimate the result of $D(\cdot)$ with minimal computational overhead. PD($\cdot$) redistributes intensity from $\supp \{r_N\}$ to $\supp \{r_P\}$ by repeated convolutions of $[-W(I|\epsilon)]_{\supp \{r_N\}}$ with the PSF until most of the intensity is transferred to $\supp \{r_P\}$  (assuming  $\supp \{r_P\}=\supp\{I\geq0\}$). We then use this approximation of $r_P$ to suppress positive ringing artifacts in the Wiener image by simply subtracting $D(-r_N\otimes h)$ from the conventional Wiener image (\textbf{Figure \ref{fig:1}C}, \textbf{Equation \ref{eq:W_NPD}}).

\begin{algorithm}[H]
  \caption{Pseudo-Deconvolution}\label{Trajectory optimisation}
  \label{al:1}
  \begin{algorithmic}[1]
    \footnotesize
    \Procedure{PD($I_{in}|k$)}{}
    \State $r_P$ = $I_{in}\cdot0$
    \State $r_N$ = $-I_{in}\cdot(I_{in}<0)$
    \State $h^T(x,y)$ = $h(-x,-y)$
    \For{$k \gets 0$ \textbf{to} $K$}
        \If{$k$ \textbf{is even}} 
        \State $I_{temp}$ = $r_N \otimes h$
        \Else
        \State $I_{temp}$ = $r_N \otimes h^T$
        \EndIf
        \State $r_P(I_{in}>0)$ += $I_{temp}(I_{in}>0)$
        \State $r_N(I_{in}<0)$ = $I_{temp}(I_{in}<0)$
        \EndFor
    \State \textbf{return} $r_P$
  \EndProcedure
\end{algorithmic}
\end{algorithm}

Similarly to MPC, negativity pseudo-deconvolution (NPD) is also capable of improving the PSNR of the PSF (\textbf{Figure \ref{fig:1}F}) and reducing $r_{eff}$ (\textbf{Figure \ref{fig:1}G}). Additionally, because the fundamental mechanisms behind MPC-Wiener and NPD-Wiener are different, the techniques can be synergistically combined to further boost their performance, as demonstrated below.

\end{multicols}
\noindent\makebox[\linewidth]{\rule{\linewidth}{0.4pt}}
\begin{equation}
    W(I|\epsilon)_{NPD}=
    \left\{\begin{matrix}
       [W(I|\epsilon)-D(-r_N\otimes h)](x,y) & \text{for } [W(I|\epsilon)-D(-r_N\otimes h)](x,y)\geq0 \\
       0 & \text{for } [W(I|\epsilon)-D(-r_N\otimes h)](x,y)<0
    \end{matrix}\right.
    \label{eq:W_NPD}
\end{equation}

\begin{equation}
    W(I|\epsilon)_{DC}=
    \left\{\begin{matrix}
       [W(I|\epsilon)-D(-r_N\otimes h)](x,y) & \text{for } \forall_{\epsilon'\geq \epsilon}:[W(I|\epsilon')-D(-r_N\otimes h)](x,y)\geq0 \\
       0 & \text{for } \exists_{\epsilon'\geq \epsilon}:[W(I|\epsilon')-D(-r_N\otimes h)](x,y)<0
    \end{matrix}\right.
    \label{eq:W_DC}
\end{equation}
\noindent\makebox[\linewidth]{\rule{\linewidth}{0.4pt}}
\begin{multicols}{2}

\section{Validation}
\begin{figure}[H]
\begin{center}
\includegraphics[width=\linewidth]{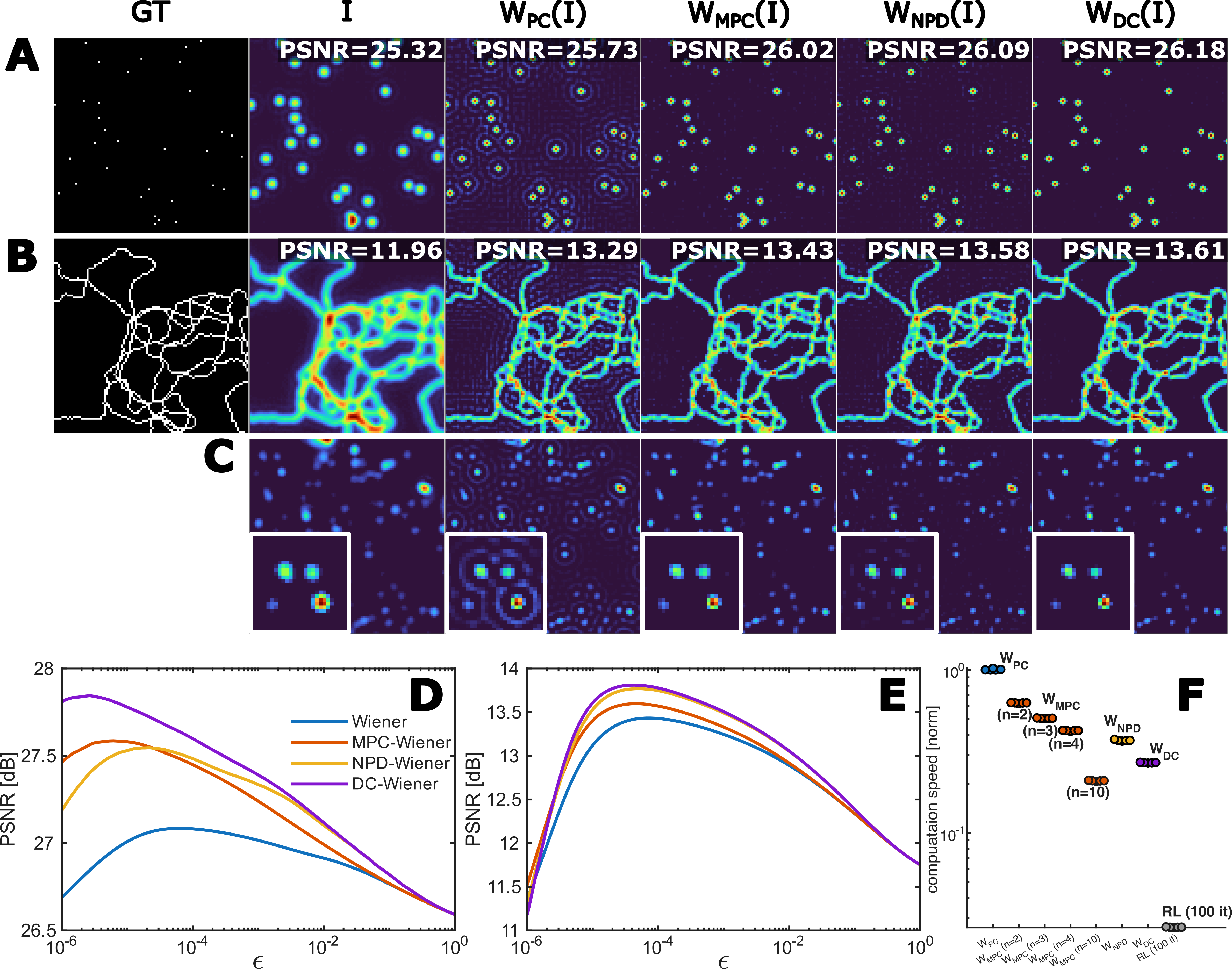} 
\caption 
{\label{fig:2}
\textbf{(A-C)} Comparison of performance between Wiener, MPC-Wiener, NPD-Wiener and DP-Wiener for sparse point \textbf{(A)} and line \textbf{(B)} computational phantoms, and for an experimentally acquired image of fluorescent beads \textbf{(C)}. \textbf{(D,E)} PSNR dependence on the regularization constant $\epsilon$ for phantoms from panels \textbf{(A,B)} respectively. \textbf{(F)} Comparison of computation times relative to conventional Wiener deconvolution.} 
\end{center}
\end{figure}

We first validate our approaches by deconvolving computationally generated phantoms of randomly distributed point sources (\textbf{Figure \ref{fig:2}A}) and filaments (\textbf{Figure \ref{fig:2}B}). The simulated objects are convolved with a PSF, adjusted to a peak intensity of $10^4 \ e^-$, and degraded by adding the corresponding amount of Poisson noise. 

We observe that in both cases regular Wiener deconvolution improves image resolution but also leads to ringing artifacts. These can be suppressed by either of our methods, leading to improvements in PSNR that can be quantified as a function of the regularization constant $\epsilon$ (\textbf{Figure \ref{fig:2}D,E}). Additionally, we observe that combining MPC-Wiener and NPD-Wiener yields further improvements in PSNR. We refer to this combined approach as double-constrained Wiener deconvolution (DC-Wiener, \textbf{Equation \ref{eq:W_DC}}).

\end{multicols}
\begin{multicols}{2}

We also validate our approaches with experimentally acquired data (\textbf{Figure \ref{fig:2}C}). Here, we use bead images acquired with a widefield microscope composed of an Olympus $20\times$ 0.75 NA objective with a Nikon 200 mm tube lens and a Thorlabs Zelux 1.4 MP camera. For deconvolution we use a Gaussian PSF model fitted to experimentally acquired bead images. Prior to deconvolution, the images are preprocessed  by removing background using the Fiji \cite{schindelin2012fiji} 'Substract background' function followed by an additional subtraction of a background estimated from a Gaussian blur of the image.

Finally, we evaluate the computation speed of our deconvolution approaches compared to conventional Wiener deconvolution (\textbf{Figure \ref{fig:2}F}). The computation speed of the fastest method (MPC$_2$-Wiener) is about 65\% that of conventional Wiener, whereas the computation speed of the slowest (DC-Wiener) is about 27\% -- still an order of magnitude faster than Richardson-Lucy (RL) deconvolution (2.7\% for 100 iterations). As can be seen for these Wiener deconvolution variants, the performance is approximately inversely related to computation speed (with DC-Wiener being both the best and the slowest), allowing the user to adjust the trade-off between performance and speed.

\section{Conclusion}

Wiener deconvolution remains one of the fastest and easiest to implement deconvolution schemes. It does however suffer from artifacts that reduce its practical usability in certain scenarios. Our work demonstrates two simple variants to Wiener deconvolution that help alleviate ringing artifacts, while remaining easy to implement and not requiring additional prior knowledge (apart from that already implicitly present within the data). The variants present a trade-off between performance and computation speed. In the case of MPC-Wiener, this trade-off is navigated by adjusting the number ($n$) of $\epsilon'$-values used for the positivity constraint. We emphasize that even small values of $n$ can provide significant benefit. That is, both variants proposed here only minimally undermine processing speed, and thus preserve one of the key advantages of Wiener deconvolution. 

\section*{Funding}
This work was funded by the National Science Foundation (EEC-1647837) and the National Institutes of Health (R01NS116139, R01GM160992).

\section*{Disclosures}
The authors declare no conflicts of interest.

\section*{Data availability} Data underlying the results presented in this paper are not publicly available at this time but may be obtained from the authors upon reasonable request. Code for performing the modified Wiener deconvolution will be available on: \url{https://github.com/biomicroscopy}.

\end{multicols}

\printbibliography

\end{document}